\documentclass[useAMS,usenatbib]{mn2e}

\def\spose#1{\hbox to 0pt{#1\hss}}
\def\simlt{\mathrel{\spose{\lower 3pt\hbox{$\mathchar"218$}}
     \raise 2.0pt\hbox{$\mathchar"13C$}}}
\def\simgt{\mathrel{\spose{\lower 3pt\hbox{$\mathchar"218$}}
     \raise 2.0pt\hbox{$\mathchar"13E$}}}

\author[Humphrey et al.]{A. Humphrey$^{1}$, F. Iwamuro$^{2}$, M. Villar-Mart\'\i n$^{3}$, L. Binette$^{1}$, R. Fosbury$^{4}$, \newauthor S. di Serego Alighieri$^{5}$ \\
$^{1}$Instituto de Astronom\'\i a, Universidad Nacional Aut\'onomo de M\'exico, Ap. 70-264, 04510 M\'exico, DF, M\'exico\\
$^{2}$Department of Astronomy, Kyoto University, Kitashirakawa, Kyoto 606-8502, Japan\\
$^{3}$Instituto de Astrof\'\i sica de Andaluc\'\i a (CSIC), Aptdo. 3004, 18080 Granada, Spain\\
$^{4}$Space Telescope - European Coordination Facility, Karl-Schwarzschild Str. 2, 85748 Garching-bei-M\"unchen, Germany\\
$^{5}$INAF-Osservatorio Astrofisico di Arcetri, Largo Enrico Fermi 5, I-50125 Firenze, Italy}
\title[]{UV and optical emission lines from the z$\sim$2.6 radio galaxy 0828+193: spatially resolved measurements}

\begin{document}

\date{Accepted 2007 September 12.  Received in original form 2007 August 17.}

\pagerange{\pageref{firstpage}--\pageref{lastpage}} \pubyear{2007}

\maketitle

\label{firstpage}

\begin{abstract}
We present an investigation into the spatial variation of the rest-frame UV and optical line and continuum emission along the radio axis of the z$\sim$2.6 radio galaxy 0828+193, using long-slit spectra from the Keck II and Subaru telescopes.  Line brightnesses, line ratios and electron temperatures are examined, and their relationship with the arm-length asymmetry of the radio source is also investigated.  We find that on the side of the nucleus with the shortest radio lobe, the gas covering factor is higher, and the ionization parameter is lower.  The contrasts in environmental density required to explain the asymmetries in the line brightness and the radio arm-length asymmetries are in fair agreement with each other.  These results add further weight to the conclusion of McCarthy, van Breugel \& Kapahi (1991) -- lobe distance asymmetries in powerful radio sources are the result of an asymmetry in the environmental density.  

We also note that the brightness of both the UV and optical continuum emission shows a similar spatial asymmetry to that shown by the line emission.  While the UV continuum asymmetry can be wholly explained by the expected asymmetry in the nebular continuum, the optical continuum asymmetry cannot.  We argue that, at least at optical wavelengths, the starlight and/or the scattered light must also be strongly spatially asymmetric.

\end{abstract}


\section[]{Introduction}

The study of powerful radio sources and their environments continues to be important for cosmological debates.  Their bright line and continuum emission allows the selection of large samples across an enormous range in redshift, and since they are thought to be hosted by massive galaxies, they can be used as beacons to massive galaxies (e.g. Seymour et al. 2007).  

Powerful radio galaxies at high redshift are often embedded within ionized nebulae which emit intense emission lines from a wide range of species, and across large spatial scales (up to $\sim$100 kpc in some cases: Villar-Mart\'\i n et al. 2002).  These haloes of ionized gas can provide information about the evolutionary status of the host galaxy and also about the possible symbiosis between the AGN related activity and the evolution of the host galaxy and its environment.  

It is now recognised that the radio jets can interact with the ambient interstellar medium (ISM hereinafter) in such a way as to perturb the kinematic properties of the gaseous haloes: split kinematic components, higher full-width at half of maximum (FWHM hereinafter) and substantial velocity shifts are often seen to be associated with hotspots and bends in the radio structures (e.g. Villar-Mart\'\i n et al. 1999).  Recent results have shown that this perturbed gas is likely to be in outflow, although it is not clear whether this outflowing gas will eventually escape the gravitational potential of the host galaxy (Humphrey et al. 2006; Nesvadba et al. 2006).  

In addition to this perturbed gas, the haloes of powerful radio sources also contain gas with quiescent kinematics, i.e., gas with FWHM and velocity shifts that are consistent with gravitational motion within/around a massive galaxy ($\la$700 km s$^{-1}$: Villar-Mart\'\i n et al. 2002, 2003).  At least at high-z, this quiescent gas appears to be in infall (Humphrey et al. 2007a), and may represent the fuel for the AGN activity and star formation.  

Ratios between emission lines can also be useful for understanding the evolutionary status of the host galaxy.  Given the right combination of emission lines, the ionization state, the ionization mechanism and the metallicity can be determined (e.g. Robinson et al. 1987).  In this paper we investigate the spatial variation in emission lines along the radio axis of the z$\sim$2.6 radio galaxy 0828+193.  We combine high-S/N spectra from Keck II and Subaru in order to examine together UV and optical emission lines that will allow us to diagnose the physical conditions, and the spatial variation thereof, in the extended emission line region (EELR hereinafter).  Previous studies have shown that considering the rest-frame UV lines together with the rest-frame optical lines is crucial for disentangling various ionization and abundance effects (see e.g. Iwamuro et al. 2003; Humphrey et al. 2007b).  Throughout this paper we assume a flat universe with $H_{0}$ = 71 km $s^{-1}$ Mpc$^{-1}$, $\Omega_{\Lambda}$ = 0.73 and $\Omega_{m}$ = 0.27.

\section[]{Observations and data analysis}

The optical spectrum (sampling the rest-frame UV emission) was obtained at the Keck II telescope during December 1997, using the Low Resolution Imaging Spectrometer (LRIS; Oke et al. 1995) in polarimetry mode (Goodrich et al. 1995), with a total integration time of 18000 s.  A 300 line mm$^{-1}$ grating was used, which resulted in a dispersion of 2.4\AA~pixel$^{-1}$ and an instrumental profile with FWHM $\sim$10.5\AA~($\sim$500 km s$^{-1}$).  This instrumental set-up provided a (rest-frame) wavelength coverage of $\sim$1200-2400\AA, i.e from Ly$\alpha$ to [NeIV] $\lambda$2424.  A 1\arcsec slit was used, which was oriented along the radio axis (44\degr; Carilli et al. 1997).  The FWHM of the seeing disk, determined from an M5V star within the slit, was $\sim$1.7\arcsec throughout the observations, and the plate scale was 0.214\arcsec pixel$^{-1}$.  See Vernet et al. (2001) and Villar-Mart\'\i n et al. (2002) for a more detailed description of the observation and reduction of this spectrum.  

The near infrared spectrum (which samples the rest-frame optical emission) was observed on 18 December 2000 at the Subaru telescope, using the OH Airglow Supressor (OHS hereinafter) spectrograph (Iwamuro et al. 2001), for a total exposure time of 4000 s in 0.6\arcsec seeing conditions.  The spatial scale is 0.111\arcsec pixel$^{-1}$ and the wavelength scale is 8.5\AA~ pixel$^{-1}$.  J and H band spectra were observed simultaneously, giving a rest-frame spectral range of $\sim$3100-3800\AA~ and $\sim$4100-5100\AA, respectively.  The spectral resolution is relatively low ($\sim$1400 km s$^{-1}$).  The slit was oriented at a position angle of 38\degr (i.e. close to the radio PA), and was 1\arcsec wide.  The observation and reduction of the Subaru spectrum is discussed in greater detail by Iwamuro et al. (2003).  

The Keck II and Subaru two-dimensional spectra were aligned spatially using the field star to the SE of the radio galaxy.  The maximum uncertainty in this alignment is $\sim$0.2\arcsec.  We assume that the peak of the NIR continuum marks the location of the active nucleus.  In addition the reduction steps described by Vernet et al. (2001), Villar-Mart\'\i n et al. (2002) and Iwamuro et al. (2003), we have performed a flux calibration that will allow us to compare directly the line fluxes measured in the Keck II and Subaru spectra.  We scaled the Keck II spectrum such that in the appropriate wavelength range, its flux measured in a 4.1\arcsec$\times$1.0\arcsec extraction aperture matches that measured from a 4.1\arcsec$\times$1.0\arcsec pseudoslit on the {\it HST} F675w image (Pentericci et al. 1999) after it was smoothed to a spatial resolution (FWHM) of 1.7\arcsec.  Since FWHM of the seeing disc was much larger than the width of the slit, we performed a correction for the expected slit loss.  The uncertainty in the absolute flux scale of the Keck spectrum is expected to be about 20 per cent.  

Similarly, the Subaru spectrum was scaled such that the total H-band flux measured in a 2.3\arcsec$\times$1.0\arcsec extraction aperture matches that measured from a 2.3\arcsec$\times$1.0\arcsec pseudo slit on the H-band image of Iwamuro et al. (2003).  The uncertainty in the resulting absolute flux scale is expected to be roughly 10 per cent.  The two-dimensional spectra are shown in Figure 1. 

Although the Keck and Subaru spectra were taken using slightly different values for PA (44\degr and 38\degr, respectively), we feel that this is not likely to have a major impact on our conclusions, since most of the line ratios used in our analysis use lines selected solely from one of the two spectra.  Where we do form line ratios from lines from both spectra (i.e. OIII] $\lambda$1663 / [OIII] $\lambda$5007), we try to minimize the possible impact of the different PA and seeing FWHM.  

One-dimensional spectra were extracted at various positions along the slit, with apertures selected in order to make the most physical sense (see Tables 1 and 2 for aperture definitions).  First, we extracted separate spectra for the NE and SW regions of the EELR from relatively large apertures (A and B; $\sim$1\arcsec in diameter), in order to obtain relatively high signal-to-noise measurements of the fainter lines [OIII] $\lambda$ 4363 and He II $\lambda$4686 for these two regions.  Then we extracted spectra from much smaller apertures along the slit (C-J for the Subaru spectra, and K-O for the Keck spectra, respectively), in order to examine the variation of physical conditions with finer spatial sampling, but at the expense of some fainter lines.  Figure 1 shows the spatially aligned 2-dimensional spectra.  Figure 2, 3 and 4 show the spatial profiles along the slit of the rest-frame optical lines, rest-frame UV lines, and the continuum emission, respectively.  These spatial profiles have had the continuum emission subtracted.  In all figures, we adopt the position of the NIR continuum peak (which also corresponds to the peak of the emission in the {\it HST} F657W image of Pentericci et al. 1999) as the spatial zero.  We used the field star to register the Keck and Subaru spectra.  

Measurements of the line fluxes were then carried out using the {\tt STARLINK DIPSO} package.  Since the line velocity profiles appear non-Gaussian at several positions along the slit in the Keck spectrum, simple volume integrals were used, rather than Gaussians.  The continuum flux density was measured from line- and residual-free regions of the spectrum, and was subtracted from the spectrum prior to measuring the line fluxes.  The formal 1$\sigma$ errors are dominated from the uncertainty in the continuum level.  Table 1 gives the line flux measurements from the Subaru spectrum, relative to H$\beta$.  Table 2 gives the fluxes measured from the Keck II spectrum, relative to HeII $\lambda$1640.  We have also measured continuum flux densities in apertures A and B.  These are listed in Table 3.  

\begin{table*}
\centering
\caption{Line measurements from the Subaru spectrum.  Columns are as follows: (1) Aperture designation; (2) position in arc seconds of the aperture along the slit, relative to the peak of the NIR continuum emission; (3) H$\beta$ surface brightness in units of 10$^{-16}$ erg s$^{-1}$ cm$^{-2}$ arcsec$^{-2}$; (4) the [NeV] $\lambda$3426/H$\beta$ ratio; (5) the [OII] $\lambda$3727/H$\beta$ ratio; (6) the H$\gamma$+[OIII] $\lambda$4363/H$\beta$ ratio; (7) the HeII $\lambda$4686/H$\beta$ ratio; (8) the [OIII] $\lambda$5007/H$\beta$ ratio; (9) the [OII] $\lambda$3727/[OIII] $\lambda$5007 ratio; (10) the [NeV] $\lambda$3426/[OIII] $\lambda$5007 ratio.} 
\begin{tabular}{llllllllll}
\hline
Ap. & Pos. & H$\beta$ & [NeV]/H$\beta$ & [OII]/H$\beta$ & H$\gamma$+[OIII]/H$\beta$ & HeII/H$\beta$ & [OIII]/H$\beta$ & [OII]/[OIII] & [NeV]/[OIII] \\
(1) & (2) & (3) & (4) & (5) & (6) & (7) & (8) & (9) & (10) \\
\hline
A & 1.17SW--0.06SW & 1.8$\pm$0.1 & 1.01$\pm$0.05 & 1.3$\pm$0.1 & 0.72$\pm$0.05 & 0.28$\pm$0.05 & 7.7$\pm$0.4 & 0.17$\pm$0.01 & 0.13$\pm$0.02\\
B & 0.06NE--1.17NE & 2.74$\pm$0.06 & 0.7$\pm$0.1 & 1.74$\pm$0.03 & 0.64$\pm$0.03 & 0.30$\pm$0.02 & 8.6$\pm$0.3 & 0.20$\pm$0.01 & 0.08$\pm$0.02\\
\hline
C & 1.17SW--0.39SW & 1.6$\pm$0.1 & 0.9$\pm$0.1 & 1.0$\pm$0.1 & - & - & 7.2$\pm$0.1 & 0.13$\pm$0.01 & 0.12$\pm$0.02 \\
D & 0.39SW--0.17SW & 1.8$\pm$0.1 & 1.1$\pm$0.1 & 1.7$\pm$0.3 & - & - & 9.5$\pm$0.2 & 0.18$\pm$0.03 & 0.12$\pm$0.01 \\
E & 0.17SW--0.06NE & 1.7$\pm$0.1 & 1.1$\pm$0.1 & 2.3$\pm$0.1 & - & - & 12.2$\pm$0.1 & 0.19$\pm$0.01 & 0.09$\pm$0.01 \\
F & 0.06NE--0.28NE & 2.7$\pm$0.2 & 0.83$\pm$0.08 & 1.58$\pm$0.07 & - & - & 8.7$\pm$0.1 & 0.18$\pm$0.01 & 0.10$\pm$0.01 \\
G & 0.28NE--0.50NE & 2.8$\pm$0.3 & 0.84$\pm$0.09 & 1.62$\pm$0.07 & - & - & 8.9$\pm$0.1 & 0.17$\pm$0.01 & 0.10$\pm$0.01 \\
H & 0.50NE--0.72NE & 2.6$\pm$0.2 & 0.65$\pm$0.07 & 1.51$\pm$0.08 & - & - & 9.4$\pm$0.1 & 0.16$\pm$0.01 & 0.07$\pm$0.01 \\
I & 0.72NE--1.39NE & 1.8$\pm$0.2 & 0.4$\pm$0.2 & 2.1$\pm$0.1 & - & - & 10.1$\pm$0.2 & 0.21$\pm$0.01 & 0.05$\pm$0.02 \\
J & 1.39NE--1.83NE & 0.6$\pm$0.1 & $\le$1.1 & 2.7$\pm$0.3 & - & - & 10.1$\pm$0.5 & 0.27$\pm$0.03 & $\le$0.1 \\
\hline
\end{tabular}
\end{table*}

\begin{table*}
\centering
\caption{Line measurements from the Keck II spectrum.  Columns are as follows: (1) Aperture designation; (2) position in arc seconds of the aperture along the slit, relative to the peak of the NIR continuum emission; (3) HeII $\lambda$1640 surface brightness in units of 10$^{-16}$ erg s$^{-1}$ cm$^{-2}$ arcsec$^{-2}$; (4) FWHM of HeII $\lambda$1640 in km s$^{-1}$, deconvolved from the instrumental profile; (5) the NV $\lambda$1240/HeII $\lambda$1640 ratio; (6) the CIV $\lambda$1550/HeII $\lambda$1640 ratio; (7) the CIII] $\lambda$1909/HeII $\lambda$1640 ratio; (8) the OIII] $\lambda$1663/HeII $\lambda$1640 ratio; (9) the CIV $\lambda$1550/CIII] $\lambda$1909 ratio; (10) the NV $\lambda$1240/CIV $\lambda$1550 ratio.} 
\begin{tabular}{llllllllll}
\hline
Ap. & Pos. & HeII & FWHM & NV/HeII & CIV/HeII & CIII]/HeII & OIII]/HeII & CIV/CIII] & NV/CIV \\  
(1) & (2) & (3) & (4) & (5) & (6) & (7) & (8) & (9) & (10)\\
\hline
A & 1.18SW--0.11SW & 3.4$\pm$0.2 & 960$\pm$40 & 0.75$\pm$0.03 & 1.83$\pm$0.08 & 0.46$\pm$0.06 & 0.28$\pm$0.02 & 4.0$\pm$0.2 & 0.41$\pm$0.02 \\
B & 0.11NE--1.18NE & 2.1$\pm$0.2 & 1280$\pm$50 & 0.45$\pm$0.07 & 1.50$\pm$0.04 & 0.53$\pm$0.05 & 0.25$\pm$0.01 & 2.8$\pm$0.8 & 0.30$\pm$0.05 \\
\hline
K & 2.27SW--1.41SW & 0.8$\pm$0.1 & 910$\pm$70 & - & 2.2$\pm$0.4 & - & 0.25$\pm$0.06 & - & - \\
L & 1.41SW--0.56SW & 2.7$\pm$0.2 & 880$\pm$40 & - & 2.4$\pm$0.1 & - & 0.21$\pm$0.02 & - & - \\
M & 0.56SW--0.30NE & 4.2$\pm$0.2 & 1240$\pm$50 & - & 1.80$\pm$0.05 & - & 0.16$\pm$0.01 & - & - \\
N & 0.30NE--1.16NE & 4.4$\pm$0.2 & 1270$\pm$50 & - & 1.56$\pm$0.06 & - & 0.15$\pm$0.01 & - & - \\
O & 1.16NE--3.94NE & 0.9$\pm$0.1 & 1190$\pm$70 & - & 1.6$\pm$0.2 & - & 0.21$\pm$0.07 & - & - \\
\hline
\end{tabular}
\end{table*}

\begin{table*}
\centering
\caption{Continuum flux densities, measured from the Keck II and Subaru spectra, and from the HST WFPC2 F675W image (Pentericci et al. 1999).  The first column gives the aperture name.  See Table 1 and 2 for details of the spatial extent of the apertures.  Columns 2-7 give continuum flux densities, in units of 10$^{-18}$ erg s$^{-1}$ cm$^{-2}$ arcsec$^{-2}$ \AA$^{-1}$.  The column headings give the wavelength for which the measurements were taken, along with the rest-frame wavelength (in brackets), and the telescope.  } 
\begin{tabular}{lllllll}
\hline
    & 4210-4540 \AA & 5600-5800 \AA & 6320-7180 \AA & 6880-6940 \AA & 12400-13200 \AA & 16600-17600 \AA \\  
    & (1180-1270 \AA) & (1570-1620 \AA) & (1770-2010 \AA) & (1920-1940 \AA) & (3470-3700 \AA) & (4650-4930 \AA) \\
Ap. & Keck II & Keck II & HST F675W & Keck II & Subaru & Subaru \\
\hline
A & 2.29$\pm$0.06 & 2.00$\pm$0.02 & 2.01$\pm$0.04 & 1.32$\pm$0.06 & 0.33$\pm$0.04 & 0.18$\pm$0.03 \\
B & 2.84$\pm$0.06 & 2.48$\pm$0.02 & 3.47$\pm$0.06 & 1.74$\pm$0.04 & 0.66$\pm$0.03 & 0.37$\pm$0.04 \\
\hline
\end{tabular}
\end{table*}

\section[]{Results}

\subsection[]{Emission lines}
In Figures 2 and 3 we show the spatial distribution along the slit of the brightest emission lines, after subtraction of the continuum emission.  Several emission lines show a striking spatial asymmetry (e.g. [OIII] $\lambda\lambda$4959,5007, CIII]), such that their surface brightness is higher on the NE side of the nucleus than on the SW side.  Other emission lines, such as [NeV] and NV, appear to be much more symmetric about the nucleus.  We parameterize the line brightness asymmetry as R, the ratio of the surface brightness in aperture B divided by the surface brightness in aperture A (Table 4).  

Interestingly, there appears to be a trend between R and ionization stage, in the sense that higher-ionization metal lines have lower values for R than do lower-ionization metal lines.  For example, the low ionization line [OII] has R=2.1, while the high-ionization lines [NeV], NV and CIV have R=1.0, 0.8 and 1.1, respectively.  The 'intermediate' ionization lines [OIII] $\lambda\lambda$4959,5007, CIII] $\lambda$1909 and OIII] $\lambda$1663 have R=1.7, 1.6 and 1.2 respectively.  We find no significant difference in R between the HeII and (non-resonant) HI recombination lines.  All the lines that show a significant spatial asymmetry have their highest surface brightness on the side of the shortest radio lobe as measured from the Very Large Array (VLA hereinafter) radio images of Carilli et al. (1997).  As shown by McCarthy, van Breugel \& Kapahi (1991), this is the case in essentially all powerful radio galaxies.   We have not considered the Ly$\alpha$ line in this analysis because its spatial asymmetry is likely to be the result of resonance scattering and orientation effects (see Humphrey et al. 2007a).  The Ly$\alpha$ asymmetry discussed by Humphrey et al. (2007a) is relative to the other emission lines, and should not be confused with the asymmetries we discuss herein.  

As a result of the differing spatial distributions of the lines, many of the line ratios show a significant variation along the slit.  In the case of the rest-frame optical line ratios, [OII]/[OIII] and [OII]/H$\beta$ increase, while [NeV]/[OIII] and [NeV]/H$\beta$ decrease, from the SW to the NE (see Table 1).  [OIII]/H$\beta$ shows no clear trend along the slit.  The HeII $\lambda$4686/H$\beta$ ratio shows no significant difference between A and B.  Turning now to the rest-frame UV lines, we find that NV/HeII, NV/CIV, CIV/HeII and CIV/CIII] are significantly higher to the SW than to the NE.  OIII]/HeII and CIII]/HeII show no significant spatial variation between A and B (Table 2).  

The fact that the HeII $\lambda$1640/HeII $\lambda$4686 ratio shows no significant difference between A and B suggests that there is no significant difference in reddening between the NE and SW EELR.  Indeed, the HeII $\lambda$1640/HeII $\lambda$4686 ratios (A:6.7$\pm$1.3 and B:5.5$\pm$0.4) are near to the theoretical case B value (7.8), suggesting that the UV-optical line spectrum is essentially unreddened (see also Humphrey et al. 2007b).  

We have calculated the electron temperature T$_{e}$ in apertures A and B, using the OIII] $\lambda$1663/ [OIII] $\lambda$5007 ratio (Humphrey et al. 2007b) and the [OIII] $\lambda$4363/ [OIII] $\lambda$5007 ratio.  For the former, we eliminated any systematic errors in the flux cross calibration (and any reddening effects) by using OIII] $\lambda$1663/ [OIII] $\lambda$5007 $\times$ 7.8 $\times$ HeII $\lambda$4686/ HeII $\lambda$1640.  Since [OIII] $\lambda$4363 is blended with H$\gamma$, we obtained the flux of the former by subtracting the H$\gamma$ flux expected from Case B recombination (i.e. H$\gamma$=0.47$\times$H$\beta$: Dopita \& Sutherland 2003).  Our use of the theoretical case B ratios for H$\gamma$/H$\beta$ is justified by the very low reddening of this source (see above).  The OIII] $\lambda$1663/ [OIII] $\lambda$5007 ratio gives T$_{e}$=16100$\pm$500 K in aperture A and T$_{e}$=15500$\pm$500 K in B; [OIII] $\lambda$4363/ [OIII] $\lambda$5007 gives T$_{e}$=16500$\pm$1800 K in aperture A and T$_{e}$=13400$\pm$1100 K in aperture B.  Both line ratios give consistent results, and we find no significant difference in T$_{e}$ between A and B.  The temperatures are given in Table 5.  

\subsection[]{Continuum}
In Figure 4 we show the spatial distribution of the rest-frame UV and optical continuum emission along the slit.  From this figure it is apparent that the continuum emission shows a similar spatial distribution and asymmetry to that shown by many of the emission lines.  In Table 6 we list the surface brightness ratio R of the continuum at various wavelengths.  R varies between 1.2 and 2.1, in good agreement with values of R measured from the emission lines.  As is the case for many of the emission lines, we find that the continuum emission is brighter on the side of the nucleus with the shortest radio lobe.  Values of R derived from the Keck II spectrum appear to be lower than those derived from the HST image and from the Subaru spectrum.  

\subsection[]{Uncertainties: seeing effects}
Before drawing conclusion from these results, it is important to consider how the difference in seeing conditions between the Keck and Subaru observations might affect our interpretation.  To examine this effect, we have smoothed the Subaru spectrum with a Gaussian filter such that the FWHM of the field star matches that in the Keck spectrum (FWHM$\sim$1.7\arcsec).  We find that the spatial asymmetry of [OIII] $\lambda\lambda$4959,5007 is reduced to R=1.4$\pm$0.1, down from R=1.7$\pm$0.1 in the unsmoothed spectrum.  In the case of the H-band continuum, we measure R=1.2$\pm$0.2, down from R=2.1$\pm$0.5.  Therefore, we conclude that the poor seeing conditions during the Keck observations are likely to have reduced the asymmetry of the rest-frame UV emission compared to the rest-frame optical emission.  In the case of the continuum, the difference in the FWHM of the seeing disc is able to explain the difference in the R values measured from the Keck and Subaru spectra.  

The difference in seeing can also explain the different values of R shown by lines from {\it the same species} (i.e., HeII; OIII).  However, this effect is unable to explain the trend for metal lines from lower ionization species to show higher values for R ($\S$3.1; Table 4) -- the trend can be clearly seen when considering the lines from the Subaru spectrum only ([OII], [OIII], [NeV]) or from the Keck spectrum only (CIII], OIII], CIV, NV).  We conclude, therefore, that this trend is real.

\section[]{Discussion}

\subsection[]{Gradient in line ratios}

We now consider several scenarios for the origin of the spatial variation (or non-variation) of the line ratios: an ionization parameter gradient; a metallicity gradient; a gradient in the impact of shock-ionization; and a gradient in the impact of stellar photoionization.  Where necessary, we use ionization model calculations as a guide for the way in which the line ratios depend on these parameters.  However, we do not attempt to {\it fit} the observed line ratios, due to the large number of free parameters in the models.  

\subsubsection[]{Ionization parameter}
In order to investigate whether the observed gradient in line ratios can be explained by a gradient in ionization parameter\footnote{We define ionization parameter U as $\frac{L}{4 \pi r^{2} n_{H} c}$, where L is the ionizing photon luminosity of the source} U, we compute photoionization models using the multipurpose code {\tt mappings} Ic (Binette, Dopita \& Tuohy 1985; Ferruit et al. 1997).  The models consist of an isobaric, plane-parallel slab of gas in the low density limit (n$_{H}$=100 cm$^{-3}$) onto which an ionizing continuum of the form {\it f$_{v}$} $\propto$ {\it v}$^{+\alpha}$, where $\alpha$=--1.5, impinges.  We adopt solar abundances (Anders \& Grevesse 1989).  We vary the ionization parameter U from 0.005 to 0.16 in steps of $\times$2.  We adopt these model parameters because they are able to give a reasonable fit to the spatially integrated line ratios of the EELR of powerful radio galaxies (e.g. Robinson et al. 1987).  

Table 7 shows results from the U-sequence calculations.  As U decreases, the effect on the optical line ratios is dramatic: [OII]/[OIII] and [OII]/H$\beta$ increase while [NeV]/[OIII] and [NeV]/H$\beta$ both decrease.  In 0828+193, these four ratios vary as expected for a gradient in U, implying lower U to the NE than to the SW.  The [OII]/[OIII] ratio implies that U decreases from $\sim$0.02 in aperture C (SW) to $\sim$0.005 in aperture J (NE), i.e. a decrease by factor $\sim$4.  This spatial variation in U is clearly not consistent with a simple r$^{-2}$ variation.  Considering the relatively large apertures A and B, we find that in A the gas has a factor of $\times$1.3 higher U than the gas in B.  If the AGN radiates equally into both ionization `cones', then the fact that U is lower to the NE implies that n$_{H}$ is higher there.  

The fact that [OIII]/H$\beta$ shows no apparent spatial variation in this source (aside from being slightly enhanced in the central aperture E\footnote{We note that the collisionally excited lines [NeV], [OII] and [OIII] are enhanced relative H$\beta$ at the position of the continuum peak (aperture E), suggesting that the higher-density nuclear narrow line region contributes to the line flux at this position.}), is also consistent with a gradient in U.  Photoionization models predict that at U$\ga$0.005, this ratio becomes insensitive to U; indeed the [OII]/[OIII] ratio implies that the EELR of 0828+193 is within this U-regime.  

The fact that NV/HeII is a factor of $\sim$2 lower in the NE EELR (aperture N) than in the SW EELR (aperture L) is consistent with U being a factor of a few lower in the NE EELR than in the SW EELR.  

Due to the turn-over at U$\ga$0.03 of the ratios formed from CIV, HeII, OIII] and CIII], two values are possible for a given U, and, therefore these ratios are less reliable than the optical ratios for estimating U (e.g. Humphrey et al. 2007b).  Nevertheless, the fact that most of these UV ratios show a substantial spatial variation is consistent with a gradient in U.  Thus, the optical and UV line ratios of 0828+193 are consistent with a negative gradient in U from SW to NE.

\subsubsection[]{Gas metallicity}
A negative gradient in metallicity across the EELR has been suggested by several authors in order to explain observed spatial gradients in line ratios in some radio galaxies and quasars (e.g. Overzier et al. 2001).  

However, in trying to explain the spatial variation in line ratios of 0828+193 we encounter several problems.  The ratios between [OII], [OIII], [NeV] and H$\beta$ are relatively insensitive to metallicity (see e.g. Humphrey et 2007b) and, therefore, the substantial variation of these ratios across the EELR of 0828+193 cannot be readily explained by a gradient in metallicity.  In addition, a gradient in metallicity is expected to result in a gradient in T$_{e}$, such that regions of lower metallicity ought to have higher T$_{e}$.  The lack of any significant difference in T$_{e}$ between A and B also appears to be incompatible with a gradient in metallicity.  

We therefore conclude that a gradient in metallicity does not provide a satisfactory explanation for the observed spatial gradient across the EELR of 0828+193.  

\subsubsection[]{Shock ionization}
As noted by Humphrey et al. (2006), the UV line kinematics of 0828+193 are relatively more perturbed on the NE side of the nucleus than on the SW side, suggesting that jet-gas interactions make a relatively stronger impact on the properties of the NE EELR.  In this context it is interesting to consider whether the observed spatial variation in line ratios are the result of stronger jet-induced shocks in the NE EELR.    

\begin{table}
\centering
\caption{Parameterizing the spatial asymmetries of the emission lines by R, the ratio of the surface brightness measured in aperture B to that measured in aperture A.  The metal lines are sorted by the ionization potential of the immediately lower ionization stage, i.e., in the case of CIV $\lambda$1549, we use the ionization potential of C$^{+2}$.  Notice the trend for R to be lower for the higher ionization metal lines.  } 
\begin{tabular}{ll}
\hline
Line & R \\  
\hline
H$\beta$ & 1.6$\pm$0.1 \\
H$\gamma$+$[$OIII$]$ $\lambda$4636 & 1.4$\pm$0.1 \\
HeII $\lambda$4686 & 1.7$\pm$0.3 \\
HeII $\lambda$1640 & 1.4$\pm$0.1 \\
\hline
$[$OII$]$ $\lambda$3727 & 2.1$\pm$0.2 \\
CIII$]$ $\lambda$1909 & 1.6$\pm$0.2 \\
$[$OIII$]$ $\lambda\lambda$4959,5007 & 1.7$\pm$0.1 \\
OIII$]$ $\lambda$1663 & 1.2$\pm$0.1 \\
CIV $\lambda$1550 & 1.1$\pm$0.1 \\
NV $\lambda$1240 & 0.8$\pm$0.1 \\
$[$NeV$]$ $\lambda$3426 & 1.0$\pm$0.1 \\
\hline
\end{tabular}
\end{table}

\begin{table}
\centering
\caption{Electron temperatures measured from OIII$]$ $\lambda$1663/ $[$OIII$]$ $\lambda$5007 and $[$OIII$]$ $\lambda$4363/ $[$OIII$]$ $\lambda$5007.  Columns: (1) Aperture; (2) Temperature diagnostic emission line ratio; (3) Value of the ratio; (4) Implied electron temperature.} 
\begin{tabular}{llll}
\hline
Ap. & Ratio & Value & T$_{e}$ \\  
(1) & (2) & (3) & (4)\\
\hline
A & OIII$]$ $\lambda$1663/ $[$OIII$]$ $\lambda$5007 & 0.059$\pm$0.006 & 16100$\pm$500 K \\
A & $[$OIII$]$ $\lambda$4363/ $[$OIII$]$ $\lambda$5007 & 0.024$\pm$0.005 & 16500$\pm$1800 K \\
B & OIII$]$ $\lambda$1663/ $[$OIII$]$ $\lambda$5007 & 0.051$\pm$0.005 & 15500$\pm$500 K \\
B & $[$OIII$]$ $\lambda$4363/ $[$OIII$]$ $\lambda$5007 & 0.015$\pm$0.003 & 13400$\pm$1100 K \\
\hline
\end{tabular}
\end{table}

\begin{table}
\centering
\caption{Quantifying the spatial asymmetry of the continuum emission.  R is the ratio of the surface brightness measured in Aperture B to that measured in aperture A.} 
\begin{tabular}{ll}
\hline
Telescope and wavelength range & R \\  
\hline
Subaru 16600-17600 \AA~ (H-band) & 2.1$\pm$0.5 \\
Subaru 12400-13200 \AA~(J-band) & 2.0$\pm$0.3 \\
Keck II 6880-6940 \AA~ & 1.3$\pm$0.1 \\
Keck II 5600-5800 \AA~ & 1.2$\pm$0.1 \\
Keck II 4210-4540 \AA~ & 1.2$\pm$0.1 \\
\hline
\end{tabular}
\end{table}

The electron temperature T$_{e}$ provides an important test for this scenario.  Since shock-ionized gas is expected to be significantly hotter than photoionized gas (e.g. Dopita \& Sutherland 1996; Villar-Mart\'\i n et al. 1999), regions with a greater contribution from shock-ionization should also have higher T$_{e}$.  However, we have found no significant difference in T$_{e}$ between the SW and NE EELR, and this suggests that the balance between shock- and photoionization does not vary significantly between the SW and NE regions.  

In addition, the [OIII] $\lambda$5007/H$\beta$ ratio is expected to be sensitive to the balance between shock-ionization and photoionization -- shock models predict [OIII] $\lambda$5007/H$\beta$ $\la$3 (e.g. Dopita \& Sutherland 1996), while AGN-photoionized gas can have values of up to $\sim$13 (see Table 7).  The high values of [OIII] $\lambda$5007/H$\beta$ ($\sim$7-12) in the SW and NE EELR suggests that both regions are dominated by photoionized gas.  Moreover, the lack of any substantial difference in [OIII] $\lambda$5007/H$\beta$ suggests that there is no substantial variation in the balance between shock-ionized gas and photoionized gas between the SW and NE regions.  

In summary, we find that a gradient in the balance between shock- and photoionization is unable to provide a natural explanation for the observed spatial variation of the emission line ratios.  

\subsubsection[]{Ionization by young stars}
In their study of 10 z$>$2 radio galaxies, Humphrey et al. (2006) also found that radio galaxies with a more luminous young stellar population also show stronger jet-gas interactions.  The precise nature of this trend is at present obscure.  Considering that the strength of jet-gas interactions appears to vary across the EELR of 0828+193, it is therefore of interest to consider whether there could also be a spatial variation in the balance between stellar and AGN photoionization.  

Line spectra emitted by stellar-photoionized nebulae differ from AGN-photoionized nebulae in three main ways.  Firstly, high excitation lines such as [NeV] and HeII should be much weaker (or absent) in stellar photoionized HII regions, due to the lack of high-energy photons emitted by the young stars.  Secondly, the electron temperature should be lower as a result of the lower mean energy of the ionizing photons.  Thirdly, as a result of the lower electron temperature, the collisionally excited lines should be significantly weaker relative to the recombination lines.  Thus, if stars make a substantial contribution to the ionization of the EELR (which appears to be the case in at least some objects: Villar-Mart\'\i n et al. 2007), and if the balance between stellar- and AGN-photoionization varies significantly along the radio axis, then T$_{e}$, HeII/H$\beta$, [NeV]/H$\beta$, [OIII]/H$\beta$ and [OII]/H$\beta$ ought to vary significantly across the EELR, and should be lower in regions of the EELR to which photoionization by young stars makes a larger contribution.  

The fact that we find no significant gradient in T$_{e}$, HeII/H$\beta$ or [OIII]/H$\beta$ between the NE and SW regions of 0828+193 argues strongly against this scenario.  Although the [OII]/H$\beta$ and [NeV]/H$\beta$ ratios do vary significantly across the EELR, they do so in a contradictory way: [NeV]/H$\beta$ is lower in the NE EELR, which would imply a relatively higher contribution from young stars; contrarywise, [OII]/H$\beta$ is higher in the NE EELR, which would imply a relatively lower contribution from young stars.  We therefore conclude that a gradient in the relative contribution from young stars is unable to provide a natural explanation for the observed spatial gradient in emission line ratios of 0828+193.

\begin{table*}
\centering
\caption{Results from the sequence in ionization parameter models.  T$_{OIII}$ is given in K.  The predicted values of T$_{OIII}$ are lower than the values measured in the EELR of 0828+193 -- with some fine tuning of the model parameters, we feel it is likely that this discrepancy can be resolved.}
\begin{tabular}{llllllll}
\hline
U &  [NeV]/H$\beta$ & [OII]/H$\beta$ & [OIII]/H$\beta$ & [OII]/[OIII] & [NeV]/[OIII] & NV/HeII & T$_{OIII}$\\
\hline
0.005 & 0.017 & 3.0 & 10.0 & 0.30 & 0.0017 & 0.0003 & 8800 \\
0.01 & 0.086 & 2.3 & 11.5 & 0.20 & 0.0075 & 0.0034 & 9400 \\
0.02 & 0.30 & 1.6 & 12.4 & 0.13 & 0.024 & 0.034 & 9900 \\
0.04 & 0.65 & 1.1 & 12.9 & 0.085 & 0.050 & 0.16 & 10200 \\
0.08 & 0.93 & 0.72 & 12.9 & 0.056 & 0.072 & 0.36 & 10500 \\
0.16 & 0.90 & 0.46 & 12.6 & 0.037 & 0.071 & 0.55 & 10500 \\
\hline
A & 1.01$\pm$0.05 & 1.3$\pm$0.1 & 7.7$\pm$0.4 & 0.17$\pm$0.01 & 0.13$\pm$0.02 & 0.75$\pm$0.03 & 16100$\pm$500 \\
B & 0.7$\pm$0.1 & 1.74$\pm$0.03 & 8.6$\pm$0.3 & 0.20$\pm$0.01 & 0.08$\pm$0.02 & 0.45$\pm$0.07 & 15500$\pm$500 \\
\hline
\end{tabular}
\end{table*}

\subsection[]{Implications for the spatially integrated spectrum}
A number of previous investigations have encountered problems in trying to explain the relative line fluxes in the spatially collapsed spectra of HzRG using photoionization models with a single value for U (e.g. McCarthy 1993; Best, R\"ottgering \& Longair 2000; Humphrey et al. 2007b).  In particular, the high-ionization lines imply U values that are several times higher than the values implied by the low-ionization lines.  As discussed by Vernet et al. (2001), the large extraction apertures that are typically used for the spatially integrated spectrum (i.e. $\sim$4\arcsec$\times$1\arcsec or $\sim$1000 kpc$^{3}$) could enclose a substantial range in gas density, which in turn could lead to a substantial range in U.  

Indeed, in the preceding sections of this paper we have concluded, using our spatially resolved measurements of the UV and optical emission lines, that there exists a substantial spatial gradient in U across the EELR of 0828+193.  Therefore, at least in the case of 0828+193, the spatially integrated spectrum does indeed sample ionized gas with a substantial range in U.  

Comparing our measurements of the optical emission lines (Table 1) against the results from our photoionization calculations (Table 7), we find that the ionization parameter U varies by a factor of $\sim$4 from aperture C (1.17\arcsec SW) to aperture J (1.83\arcsec NE).  This range in U is in agreement with that implied by the discrepancy between the high- and low-ionization lines in the spatially integrated spectrum (Vernet et al. 2001; Iwamuro et al. 2003), i.e., the U value implied by [NeV]/H$\beta$ is $\sim$4 times higher than the value implied by [OII]/H$\beta$.  

It is interesting to note that in each of the small extraction apertures we have used for the optical spectrum (C-J), the high-ionization lines still imply a higher U than do the low-ionization lines (Table 1 and Table 7).  It is unclear whether this is due to (i) a persistence down to small spatial scales ($\le$6 kpc or 0.7\arcsec) of the inhomogeneity in U, or (ii) a range in U along our line of sight through the EELR.  Integral field spectroscopy could be useful for resolving this issue, in the sense that it allows the variation in physical conditions away from the radio axis to be investigated (e.g. Villar-Mart\'\i n et al. 2006; Nesvadba et al. 2006).

\subsection[]{The connection between the emission line and radio source asymmetries}
We have found that many of the emission lines of 0828+193 are asymmetrically distributed along the slit (see $\S$3.1 and Figure 2).  [OIII] $\lambda$5007, H$\beta$, [OII] $\lambda$3727, HeII $\lambda$1640, HeII $\lambda$4686, and CIII] $\lambda$1909 are all clearly brighter to the NE of the nucleus than to the SW (Table 4).  We now investigate the nature of the line brightness asymmetry.  

A comparison between the measured values of HeII $\lambda$ 4686/H$\beta$ ($\sim$0.3) and the values predicted by ionization-bounded photionization models ($\alpha$=-1.5, U=0.05 results in HeII $\lambda$ 4686/H$\beta$ = 0.26) shows that the photoionized clouds in the EELR of 0828+193 are likely to be ionization-bounded.  (If the clouds were matter-bounded, then HeII $\lambda$ 4686/H$\beta$ would be significantly higher.)

In the ionization-bounded case an asymmetry in the covering factor, f$_{c}$, of these clouds (the fraction of the sky, as seen from the AGN, that is covered by the clouds) represents the most plausible way to produce an asymmetry in the brightness of H$\beta$ (assuming the AGN radiates equally into the NE ans SW `cones').  This is because the flux of H$\beta$ is proportional to the number of ionizing photons that are absorbed by the gas clouds. The H$\beta$ flux is expected to be independent of U, metallicity and n$_{H}$ (in the ionization bounded case).  The value R=1.6$\pm$0.1 we have measured from H$\beta$ thus implies that f$_{c}$ is $\sim$1.6 times higher in the NE EELR than in the SE EELR.  Since we are considering ionization-bounded clouds, the mass of ionized gas is $M_{H^{+}} \propto$ f$_{c} \times$ U.  Thus, if f$_{c}$ is 1.6 times higher in B than in A, and U is 1.3 times lower in B than in A ($\S$4.1.1), then $M_{H^{+}}$ should be $\sim$1.2 higher in B than in A.  

We now consider the relationship between the ionized gas and the radio source.  In $\S$4.1 we found that the line emission is brightest on the side of the shortest radio lobe.  As demonstrated by McCarthy, van Breugel \& Kapahi (1991), this is true for essentially all powerful radio galaxies at intermediate and high-z.  These authors proposed that the correlated line brightness and radio arm-length asymmetries arise due to inhomogeneities in the density and distribution of dense clouds: the radio source propagates more slowly, and the line emission is brighter, on the side of the nucleus with the denser ISM/IGM.  We suggest that this is also the case in 0828+193.  

We now examine whether the asymmetry in the mass of gas implied by the radio arm-length asymmetry Q of the radio source (the distance from the nucleus to the farthest radio hotspot divided by the distance from the nucleus to the closest hotspot) is consistent with the ionized gas mass asymmetry we have derived from H$\beta$.  Using the 8 GHz radio image of Carilli et al. (1997), we measure a radio arm-length asymmetry Q of 1.2.  Since Q is expected to depend on the gas mass asymmetry as $(M_{B}/M_{A})^{1/4}$ (Swarup \& Banhatti 1981), the observed Q of 1.2 implies $M_{B}/M_{A}\sim$2.1.  Although this is rather larger than the asymmetry in ionized gas mass of $\sim$1.2 implied by the emission lines, we nevertheless find it encouraging that we obtain an ionized gas mass asymmetry of the correct sign, i.e., the mass is higher on the side of the nucleus with the shortest radio lobe.  

As mentioned earlier, the ionized gas in the EELR of 0828+193 has lower U on the side of the nucleus with the shortest radio lobe (i.e. the NE side).  If the AGN radiates equally into the NE and SW `cones', the difference in U then implies higher gas densities n$_{H}$ on the side with the shortest radio lobe.  It is not clear how this asymmetry in n$_{H}$ relates to either the covering factor asymmetry or the radio arm-length asymmetry.  We note that, in principal, it would be possible for n$_{H}$ to be lower on the side with the brightest line emission or shortest radio lobe (i.e. the converse of what we measure in 0828+193).

\subsection[]{Continuum brightness asymmetry}

As described in $\S$3.2, we find that the UV and optical continuum emission shows a similar surface brightness asymmetry to the line emission (R=1.2-2.1: Table 6), and that it is brightest on the side of the nucleus with the shortest radio lobe.  (Note that the difference in R between the rest-frame UV and rest-frame optical continuum is simply an effect of the different spatial resolutions of the Keck and Subaru data: $\S$3.3.)  

An asymmetry in the line brightness will naturally lead to an asymmetry in the brightness of the nebular continuum, and we now examine whether this can explain the observed continuum asymmetries.  We have calculated the contribution made by the nebular continuum to the overall continuum fluxes in the wavelength ranges 5600-5800 \AA, 12400-13200 \AA~ and 16600-17600 \AA.  We use coefficients tabulated for T$_{e}$=10,000 K by Aller (1987), and we normalize to the observed flux of H$\beta$.  Our calculations imply that nebular continuum contributes $\sim$35, $\sim$50 and $\sim$10 per cent in the 1570-1620, 3470-3700 and 4650-4930 \AA~ bins, respectively.  

We now calculate the overall continuum asymmetry that would result if the nebular continuum has the same asymmetry as the line emission, while both the scattered light and the starlight are {\it symmetrical}.  (In order to account for the different seeing FWHM during the Keck and Subaru observations, we assume that at UV wavelengths the nebular continuum has the same asymmetry as HeII $\lambda$1640, R=1.4, and that at optical wavelengths it has the same asymmetry as H$\beta$, R=1.6).  In the 1570-1620 \AA~ bin, the overall continuum asymmetry would be R=1.2, in good agreement with the observed value (R=1.2$\pm$0.1).  In the 3470-3700 and 4650-4930 \AA~ bins, the overall R would be 1.3 and 1.1, respectively -- much lower than observed (R=2.0$\pm$0.3 and 2.1$\pm$0.5).  Therefore, the UV continuum asymmetry can be wholly explained by an asymmetry in the nebular continuum, while the optical continuum asymmetry cannot.  In order to reproduce the observed asymmetry in the optical continuum, we feel that the scattered light and/or the starlight must also be spatially asymmetric.  

Although at present we are unable to provide a detailed explanation for the asymmetry in the scattered and/or stellar light, it seems plausible that this asymmetry is related to the covering factor asymmetry of the ionized gas ($\S$4.3).  If dust is mixed with this gas, then the covering factor of the dust might also be higher on the NE side of the nucleus.  If this is the case, then we would expect a higher fraction of the quasar continuum to be scattered, and hence the scattered light to appear more luminous, on the NE side of the nucleus.  The scattering angle can also affect the brightness of scattered light, in the sense that forward scattering is more efficient than backward scattering.

In addition, if there is a larger mass of gas on the NE side of the nucleus (as implied by the higher covering factor and density: $\S$4.3), then a higher rate of star formation could result, provided that the star formation efficiency does not vary greatly between the NE and SW regions.  In a future paper, we will examine in greater detail the nature of the UV-optical continuum asymmetries shown by HzRG.  

\section[]{Conclusions}
We have presented an investigation into the spatial variation of the rest-frame UV and optical line and continuum emission along the radio axis of 0828+193, using long-slit spectra from the Keck II and Subaru telescopes.  Line brightnesses, line ratios and electron temperatures have been examined, and their relationship with the arm-length asymmetry of the radio source has been considered.  We have found that on the side of the nucleus with the shortest radio lobe, the gas covering factor is higher, and the ionization parameter is lower.  The contrasts in gas mass required to explain the line brightness and radio arm-length asymmetries are in fair agreement.  These results add further weight to the conclusion of McCarthy, van Breugel \& Kapahi (1991) -- lobe distance asymmetries in powerful radio sources are the result of an asymmetry in the environmental density.  

We have also found that the brightness of the UV and optical continuum emission shows a similar spatial asymmetry to that shown by the line emission.  While the UV continuum asymmetry can be wholly explained by the expected asymmetry of the nebular continuum, the optical continuum asymmetry cannot.  We have argued that in 0828+193, the starlight and/or the scattered light must also be strongly spatially asymmetric.  

\section*{Acknowledgments}
AH acknowledges a postdoctoral fellowship from the Universidad Nacional Aut\'onomo de M\'exico.  The work of MVM has been supported by the Spanish Ministerio
de Educaci\'on y Ciencia and the Junta de Andaluc\'\i a through the grants AYA2004-02703 and TIC-114.  LB acknowledges CONACYT through grant J-50296.  RF is affiliated to the Research and Science Support Division of the European Space Agency.  We would like to acknowledge the important contributions that Jo\"el Vernet, Marshall Cohen and Kentaro Motohara have made to this research.  We thank Chris Carilli and Laura Pentericci for providing us with radio maps of 0828+193.  Some of the data presented herein were obtained at the W.M. Keck Observatory, which is operated as a scientific partnership among the California Institute of Technology, the University of California and the National Aeronautics and Space Administration; the Observatory was made possible by the generous financial support of the W.M. Keck Foundation.  Some of the data presented herein were collected at Subaru Telescope, which is operated by the National Astronomical Observatory of Japan.

\begin{figure}
\includegraphics{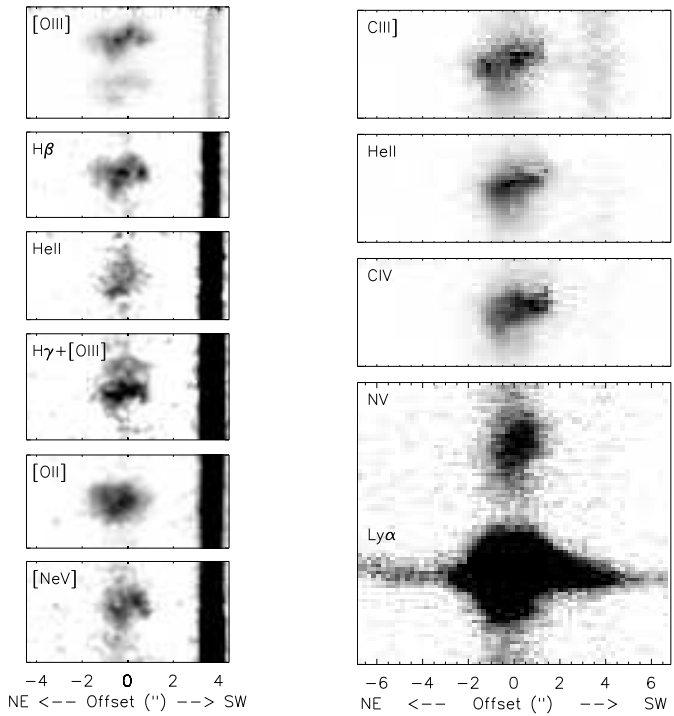} \vspace{4.0in}\caption{Sections of the two-dimensional spectra of 0828+193, showing the main emission lines.  The left column shows the two-dimensional spectrum of the optical lines [OIII] $\lambda\lambda$4959,5007, H$\beta$, HeII $\lambda$4686, the H$\gamma$+[OIII] $\lambda$4363 blend, [OII] $\lambda$3727, and [NeV] $\lambda$3426.  The right column shows the two-dimensional spectrum of the UV lines CIII] $\lambda$1909, HeII $\lambda$1640, CIV $\lambda$1550, NV $\lambda$1240, and Ly$\alpha$.  In each of the spectra, wavelength increases towards the top.   The spatial zero corresponds to the peak of the NIR (rest-frame optical) continuum emission.  The bright continuum source $\sim$4\arcsec to the SW of the radio galaxy is an M5V star (See Vernet et al. 2001).} 
\end{figure}

\begin{figure}
\includegraphics{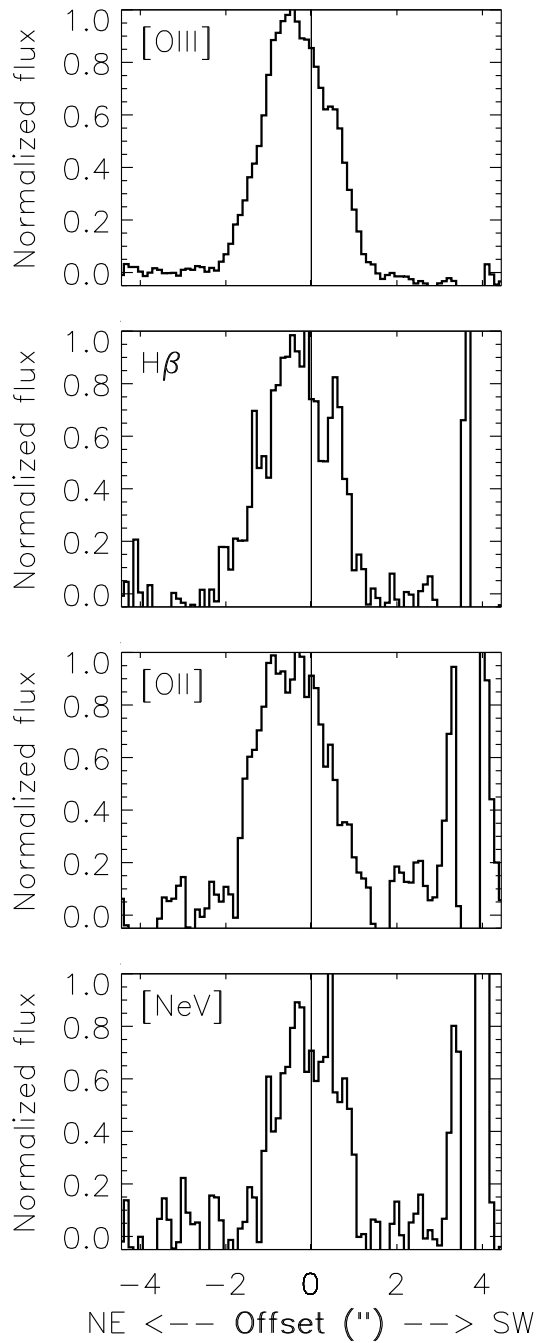} \vspace{7.3in}\caption{Spatial distribution along the slit of [OIII] $\lambda\lambda$4959,5007, H$\beta$, [OII] $\lambda$3727, and [NeV] $\lambda$3426.  The spatial zero corresponds to the peak of the NIR (rest-frame optical) continuum emission.  The spikes seen in several profiles $\sim$4\arcsec to the SW of the radio galaxy are residuals from the continuum subtraction.  Notice the trend for [OIII] $\lambda\lambda$4959,5007, H$\beta$ and [OII] $\lambda$3727 to be brightest on the NE side of the nucleus, while [NeV] $\lambda$3426 has a much more symmetrical distribution.  } 
\end{figure}

\begin{figure}
\includegraphics{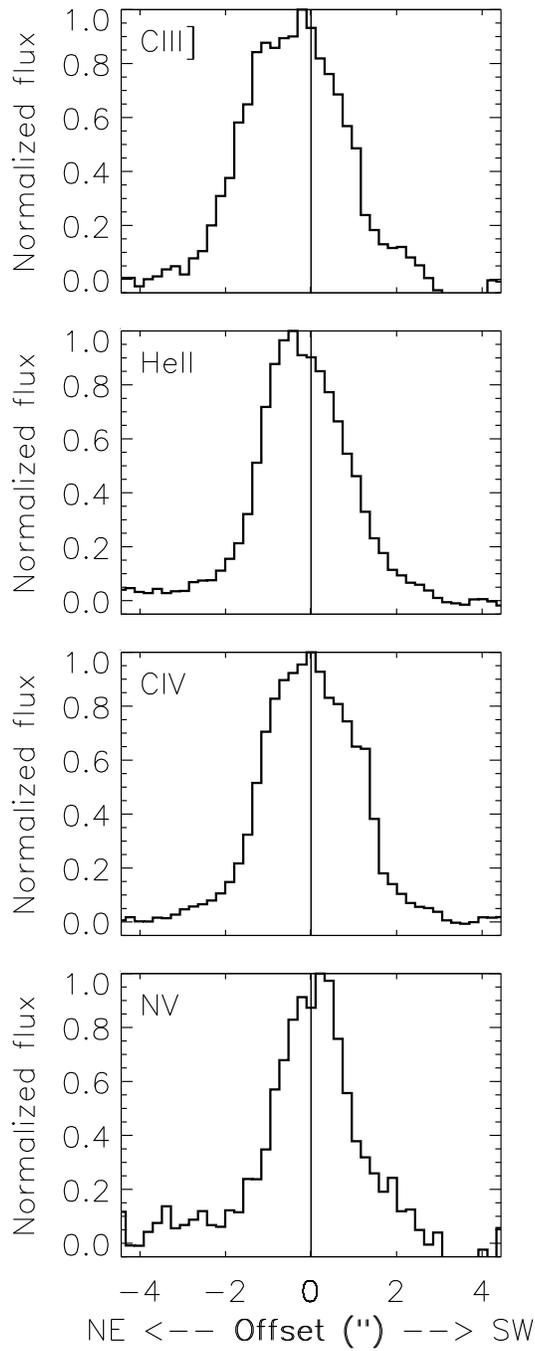} \vspace{7.3in}\caption{Spatial distribution along the slit of CIII] $\lambda$1909, HeII $\lambda$1640, CIV $\lambda$1550 and NV $\lambda$1240.  In all panels, the spatial zero corresponds to the peak of the NIR (rest-frame optical) continuum emission.  Notice the trend for CIII] $\lambda$1909 and HeII $\lambda$1640 to be brightest on the NE side of the nucleus, while CIV $\lambda$1550 and NV $\lambda$1240 are much more symmetrically distributed.} 
\end{figure}

\begin{figure}
\includegraphics{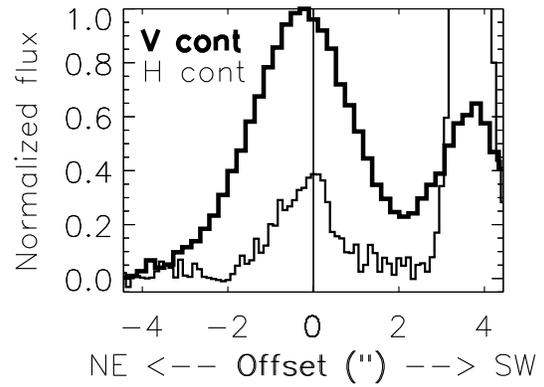} \vspace{2.2in}\caption{Spatial distribution along the slit of the rest-frame optical continuum (H-band) and rest-frame UV continuum ($\sim$V-band).  The spatial zero corresponds to the peak of the NIR (rest-frame optical) continuum emission.  As is the case for many of the emission lines, the continuum emission is brighter on the NE side of the nucleus than on the SW side.  The bright unresolved source $\sim$4\arcsec to the SW of the radio galaxy is an M5V star (Vernet et al. 2001).} 
\end{figure}

\end{document}